\documentstyle[prb,preprint,aps]{revtex}
\begin{document}
\draft
\title{Ground-state Properties of the One-dimensional Kondo Lattice
at Partial Band-filling}
\author{S. Moukouri and L. G. Caron}
\address{Centre de recherche en physique du solide and d\'epartement de
physique,\\
Universit\'e de Sherbrooke, Sherbrooke, Qu\'ebec, Canada J1K 2R1}
\maketitle

\begin{abstract}
We compute the magnetic structure factor, the singlet correlation function
and the momentum distribution of the one-dimensional Kondo lattice model at
the density $\rho =0.7$. The density matrix-renormalization group method is
used. We show that in the weak-coupling regime, the ground state is
paramagnetic. We argue that a Luttinger liquid description of the model in
this region is consistent with our calculations
. In the strong-coupling regime, the ground state
becomes ferromagnetic. The conduction electrons
show a  spinless-fermion like behavior.
\end{abstract}


The heavy-fermion materials display a variety of low temperature-states\cite
{stewart}. In some compounds, the Fermi-liquid state with huge quasiparticle
mass is stable down to the lowest attainable temperatures. In others, the
Fermi-liquid becomes magnetic or superconductor or both. The
magnetic structures vary from simple N\'eel antiferromagnetism through
incommensurate order to ferromagnetism. This rich phenomenology arises from
the interplay between a wide conduction-electron $d$-band and a partially
filled narrow $f$-band of rare-earth actinide or lanthanide elements. One
of the canonical models for the description of these systems is the Kondo
lattice model $(KLM)$.

After almost two decades of intense studies, the $KLM$
is far from being completly understood\cite{doniach}.
Even for the simplest case of one dimension, a consensus is reached only for
the half-filled case\cite{tsunetsugu1}. The model is an insulator. A spin
gap and a charge gap open for all non-zero values of the Kondo coupling. In
the metallic phase, Monte carlo\cite{troyer} and exact diagonalization\cite
{tsunetsugu2} calculations have found that the $KLM$ presents two phases. A
paramagnetic $(PM)$ state with $RKKY$ correlations is stable at low values
of the Kondo coupling. The $PM$ state evolves to a ferromagnetic $(FM)$
ground state when
the coupling becomes greater than a critical value. This $FM$ phase is found
 to be numerically in accord with the results of strong-coupling expansion%
\cite{sigrist1}. At very low densities, the model has a $FM$ phase at all
non-zero couplings\cite{sigrist2}. The Monte Carlo study was, however,
performed at finite temperature while the exact diagonalization one
was restricted
to a lattice of 8 sites. Therefore, the trends displayed in these computations
 remain to be confirmed in the ground state of much longer chains
. Using standard
notations the $KLM$ may be written as follows:

\begin{equation}
H=-t\sum_{is}(c_{is}^{+}c_{i+1s}^{}+h.c.)+J_K\sum_i{\bf S}_{ic}^{}\cdot {\bf %
S}_{if}^{}  \eqnum{1}
\end{equation}
where

\begin{equation}
{\bf S}_{ic}^\alpha =\frac 12\sum_{s,s^{\prime }}c_{is}^{+}\sigma
_{ss^{\prime }}^\alpha c_{is^{\prime }}^{},{\bf S}_{if}^\alpha =\frac 12%
\sum_{s,s^{\prime }}f_{is}^{+}\sigma _{ss^{\prime }}^\alpha f_{is^{\prime
}}^{}  \eqnum{2}
\end{equation}
with the constraint that

\begin{equation}
\sum_sf_{is}^{+}f_{is}^{}=1  \eqnum{3}
\end{equation}
the letters $c$ and $f$ stand for conduction and localized electrons
respectively. We applied the density-matrix renormalization
group $(DMRG)$ technique to this one-dimensional Kondo lattice.
We chose a density $\rho =0.7$,
 typical of those found in 3D compounds
. We believe that
the properties displayed at this density will reflect the behavior of the
model in the whole region of moderate doping. Although the physical range of
$J_K$ corresponds to small couplings, it is also interesting to study the
strong-coupling regime in order to understand the behavior of the model in
the whole range of parameters. We varied $J_K$ from $0.25$ to $10$. In the $%
DMRG$, an iteration of the algorithm consists in adding two sites at each
step. It can be immediately realized that there is a problem in keeping the
electron density fixed during the iteration process. To get around such a
problem, we constructed the reduced density matrix from the two states whose
electron numbers bracket the desired density. In the $DMRG$ method, the
states are also labelled by the $z$-component of the total spin. In the
present study, we work in the subspaces having $S_T^z=0$ $and$ $\pm \frac 12$%
. For more details, the reader is referred to a recent paper by Chen and one of
us\cite{chen}, where the method has been succesfully checked for the
one-dimensional $t-J$ model. The maximum lattice
size we have reached is 75 with up to 180 states kept in the two external
blocks. The truncation error of the Hilbert space at each iteration
is around $7. 10^{-4}$ at $J_K=0.5$ and less than
$10^{-6}$ at $J_K=10$.
We have computed the binding energy, the magnetic structure factor,
the on-site conduction electron-localized spin correlation and the electron
momentum distribution. Our results confirm the conclusions of
Monte Carlo\cite
{troyer} and exact diagonalization
\cite{tsunetsugu2} studies.
 A paramagnetic state is stable in the small
coupling regime. This state is characterized by a maximum in magnetic
structure factor at $2k_{F_c}=\pi \rho $ and a singularity in the electron
momentum distribution function at $k_{F_c}$. In the strong-coupling regime,
the ground state is $FM$. The singularity of the
conduction-electron momentum distribution
is shifted to $2k_{F_c}$.

When calculating the ground-state energy, we have taken the average of the
lowest states having $\rho_1$ and $\rho _2$ such that $\rho _1\leq \rho
\leq \rho _2$ (see Ref.9). This way,
 the ground-state energy per site of the non-interacting system can be
reproduced up to four digits. In the $J_K=\infty $ case, the $N_c$
conduction electrons form perfect on-site spin-singlets with the localized
spins. The other $N-N_c$ $f$-spins remain free. The system presents a $%
2^{(N-N_c)}$ fold spin degeneracy. When the coupling is strong but finite,
the ground-state energy per site is very close to $-\frac 34J_K\rho .$ The
binding energy is defined as $E_B=\left[ E_G(J_K=0,N)-E_G(J_K,N)\right] /N,$
where $E_G(J_K,N)$ is the ground-state energy. At a given density $\rho $, $%
E_B$ per site will be very close to $E_\infty =\frac 34\rho J_K+e_0.$ The
quantity $e_0=-\frac 4\pi \sin (\frac \pi 2\rho )$ is the energy per site of
the non-interacting case. In $Fig.1$, we show that our results are
consistent with such an analysis. The convergence to $E_\infty $ is very
smooth. However $Fig.1$ also indicates that this picture breaks down around $%
J_K=2$. Below this value of the coupling, the system enters in the small
coupling regime characterized by $\rho _FJ_K<1,\rho _F$ being the density of
state at the Fermi level of the non interacting Hamiltonian. For the
one-impurity Kondo problem, the binding energy is given by $E_1=3t(\rho
_FJ_K)^2Ln(2t)+4t\exp -\frac 1{\rho _FJ_K}$\cite{kondo}$.$ The non analytic
part of $E_1$ defines the Kondo temperature $T_K$. In the inset of $Fig.1$,
we have compared $E_B$ with $E_1$. One can see that $E_B$ is greater than $%
E_1$. This enhancement of the binding energy of the lattice over the
one-impurity case results from the intersite magnetic interaction.
However, the latter conclusion does not necessarily mean that the
Kondo temperature of the lattice is greater than $T_K$.

The magnetic properties of the $KLM$ are studied by calculating the spin
structure factors of the conduction electrons, $S_c(k$) and of localized
spins, $S_f(k)$. These quantities are defined here as follows:

\begin{equation}
S_{c,f}(k)=\frac 1N\sum_{l,m}\left\langle {\bf S}_{l,c,f}{\bf .S}%
_{m,c,f}\right\rangle\exp[i(l-m)k] .  \eqnum{4}
\end{equation}
At the first steps of the algorithm, $\rho _1$ and $\rho _2$ are
significantly different from $\rho $ and boundary effects are non
negligible. Thus, we started the calculation of the correlation
functions when the lattice size was around $35$. We have noticed that the
value of the correlation functions are extremely sensitive to that of $S_T^z$%
. We have not taken the average as for the ground-state
energy because when the
lowest state corresponding to $\rho _1$ has $S_T^z=0$, that of $\rho _2$ is $%
S_T^z=\pm \frac 12$, and $vice$ $versa$. We have used only the state with $%
S_T^z=0$. In order to reduce the effects due to the variation of the density
during the iterations, we have started the calculation of the correlation
functions at different lattice size around $35$ and then taken the average.
In agreement with previous studies, we have found that the structure factor
shows the competition of $PM$ and $FM$ phases. For small $J_K$, both $S_f(k)$
$(Fig.2)$ and $S_c(k)$ $(Fig.3)$ have a maximum at $2k_{F_c}$. The strong
spin correlation observed in $S_f(k)$ is due to the $RKKY$ interaction. As
the coupling is increased, this maximum flattens out. It completly
disappears around $J_K=2$. We note that this result compares well with that
of exact diagonalization and Monte Carlo studies. When $J_K$ is greater than
this value, a new maximum arises at $k=0$. It has been shown that in the
strong-coupling regime, the effective interaction between the $f$-spins is
ferromagnetic\cite{sigrist1}. The maximum at $k=0$ of $S_c(k)$ can be seen
as a consequence of small ferromagnetic correlations tracking
those of the $f$-spins.

The suppression of the $RKKY$ correlations can be understood by the Kondo
mechanism. When $J_K$ is small, the on-site singlet correlation $%
\left\langle {\bf S}_{i,c}{\bf .S}_{i,f}\right\rangle $ shown in $Fig.4$, is
different from that of perfect on-site singlets $-\frac 34\rho $.
This is due to the non-local character of the Kondo singlets.
 This means
that the singlet formed by an impurity and the conduction electrons has
a spatial extension. Consequently, the RKKY interaction between impurities
is favoured. As the coupling is increased, the size of the singlet reduces.
It rapidly approaches that of perfect on-site singlets. As a result, the $%
RKKY$ mechanism is suppressed. However contrary to
the conventional view of the $KLM$\cite
{fazekas}, a global singlet state is unstable against a FM state in the strong
$J_K$ limit. In this limit, the local Kondo singlets are mobile. It is the
undelying electron motion in reduced dimension that is responsible of the
$FM$ correlations.

Now, we turn to the discussion of the conduction electron momentum
distribution $n(k)$. It has been argued that in the weak-coupling regime,
the $KLM$ may present a $PM$ Luttinger liquid $(LL)$ structure\cite
{tsunetsugu2}. The $LL$ behavior is characteristic of many one-dimensional
interacting electron systems\cite{haldane}. In the $LL$ theory of $PM$
systems, $S(k)$ has a maximum at $2k_{F_c}$ and $n(k)$ presents a
singularity at $k_{F_c}$. The electron momentum distribution has the form
that follows:

\begin{equation}
n(k)=n_{k_F}-C\left| k-k_F\right| ^\alpha sgn(k-k_F)  \eqnum{5}
\end{equation}
We cannot calculate accurately the value of the exponent $\alpha $. But $%
Fig.5$ clearly shows that the singularity is located at $k=k_{F_c}$ in the
weak-coupling regime. The existence of this singularity is consistent with
the presence of the maximum of $S_{c,f}(k)$. So, the description of the
model in the weak-coupling regime in terms of $LL$ is plausible. Further
investigations of the excitation spectrum are necessary to get the full
answer. At intermediate $J_K,$ $n(k)$ is found to be very smooth so that it
becomes hard to identify any singularity. At strong $J_K$ however, the
singularity now appears at $k=2k_{F_c}$ as in a spinless fermion system.
 We believe that this is due to the
action of the Kondo coupling which supress a double occupancy of the
conduction electrons. It freezes the electron-spin degrees of
freedom. The
transition from a $PM$ to a $FM$ phase does not necessary mean that the $LL$
description breaks down. It can be yet interpreted as a transition from
a $PM$ $LL$ to a ferromagnetic $LL$.

Finally, we touch upon the question of the size of the Fermi surface in the $%
PM$ phase. In the weak-coupling region, the $KLM$ is an effective model of
the periodic Anderson model $(PAM)$. In the $PAM$ the $f$ electrons are mixed
to the conduction electron through an hybridization term. The $PAM$ is
believed to have a large Fermi surface containing both conduction electrons
and localized electrons. In the $KLM$, however, there is not any
hybridization between the two kind of particles. It is still a matter of
debate wether or not the $KLM$ has a large Fermi surface. A large Fermi
surface supposes the existence of a maximum at $2k_{F_c}+\pi $ in the
structure factor or of a singularity
 in the momentum distribution function at $k_{F_c}+%
\frac \pi 2$. Our results do not show any significant feature at this wave
number. There is yet the possibility that the singularity at the position of
the large Fermi surface is very small\cite{ueda}.

In conclusion, we have used the $DMRG$ to study the $KLM$ at the density $%
\rho =0.7.$ We believe that the behavior of the model at this density is
characteristic of the moderate doping region. In agreement with previous
exact diagonalization and Monte Carlo calculations, we have shown that the
model presents a transition around $J_K=2$. The weak-coupling region is $PM$%
. The magnetic structure factor has a maximum at $2k_{F_c}$. The electron
momentum distribution function displays a singularity at $k_{F_c}$. We have
argued that this is consistent with the description of the model in terms of
$LL$. The strong-coupling region is $FM$. In this phase, the singularity in
the electron momentum distribution is shifted to $2k_{F_c}$. This can be
interpreted as the presence of a ferromagnetic $LL$ structure.

We wish to thank Liang Chen for useful discussions. This work was supported
by a grant from the Natural Sciences and Engineering research Council
(NSERC) of Canada and the Fonds pour la formation de Chercheurs et d'Aide
\`a la Recherche (FCAR)of the Qu\'ebec government.

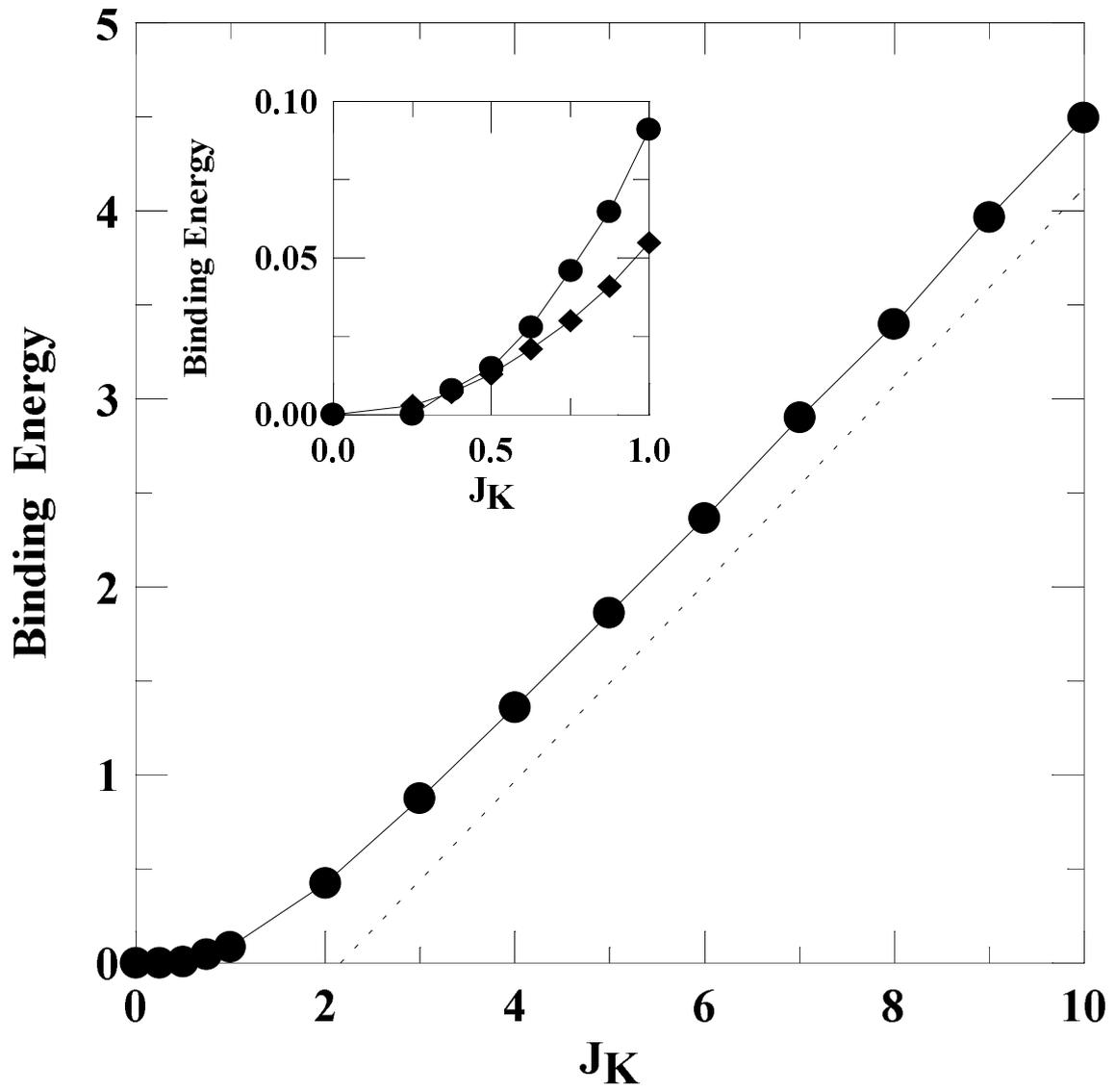
\begin{figure}[tbp]
\caption{ The binding energy of the Kondo lattice $E_B$ (circles) versus the
Kondo coupling.
 The dashed line corresponds to the strong-coupling limit.
In the inset $E_B$ is compared with the binding energy of
the one-impurity (diamonds) Kondo problem.}
\end{figure}
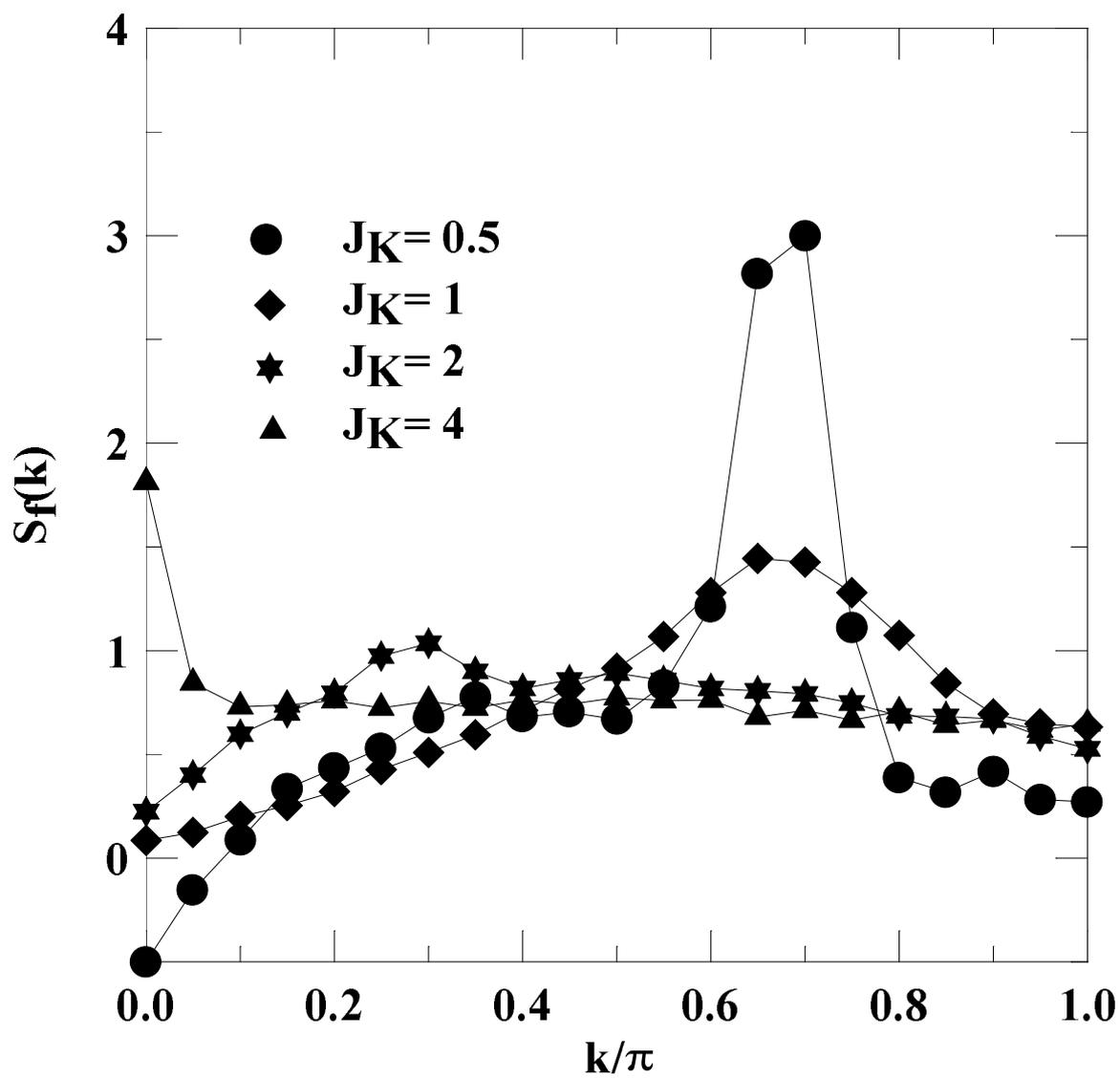
\begin{figure}[tbp]
\caption{ The magnetic structure factor of the localized spins for $\rho
=0.7 $ at $J_K=0.5$ (circles), $J_K=1$ (diamonds), $J_K=2$ (stars) and $J_K=4
$ (triangles).}
\end{figure}
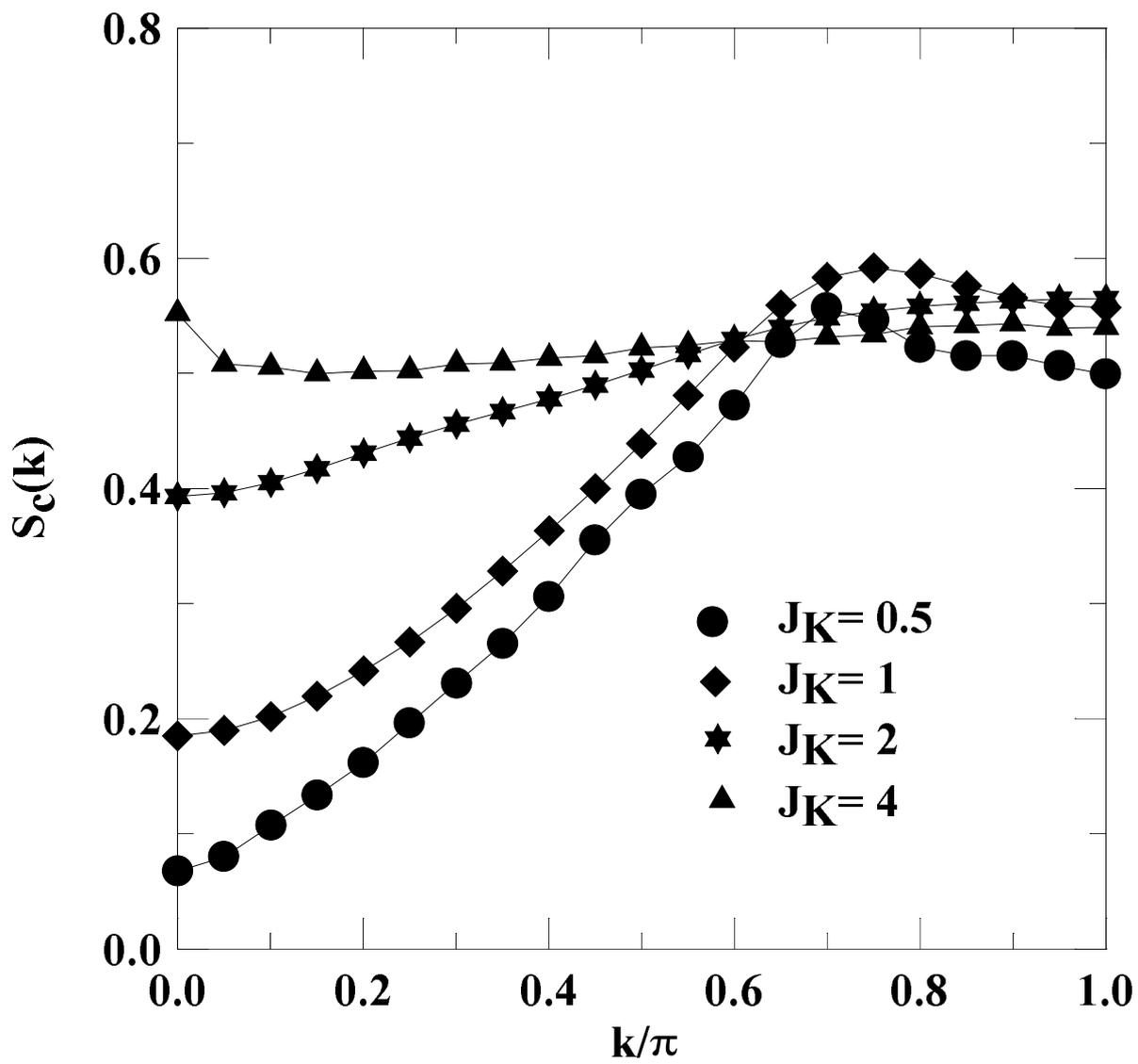
\begin{figure}[tbp]
\caption{ The magnetic structure factor of the conduction electrons for $%
\rho =0.7$ at $J_K=0.5$ (circles), $J_K=1$ (diamonds), $J_K=2$ (stars) and $%
J_K=4$ (triangles).}
\end{figure}
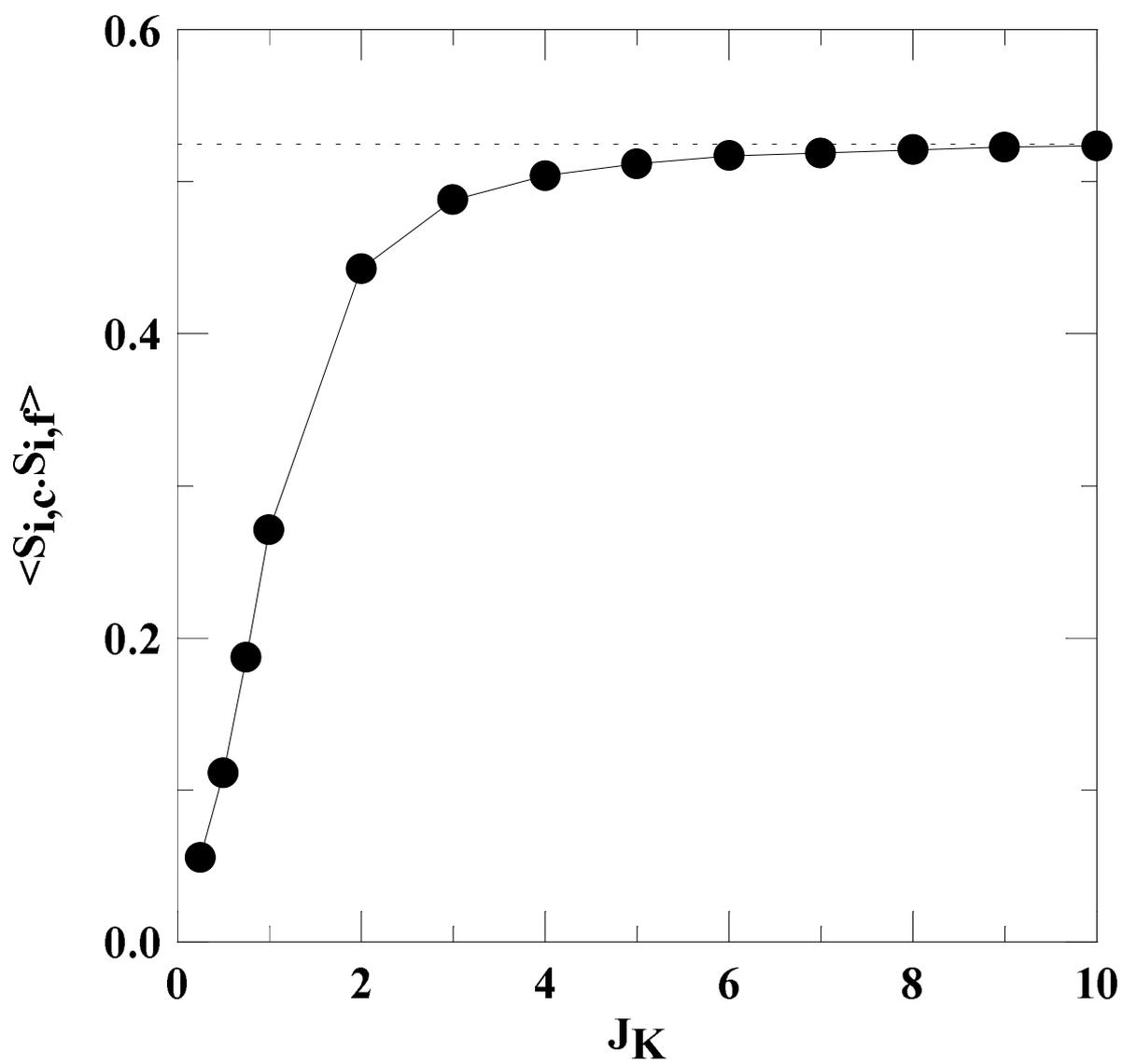
\begin{figure}[tbp]
\caption{ The on-site conduction electron-localized spin correlation versus
the Kondo coupling. The dashed line corresponds to the strong-coupling
 limit}
\end{figure}
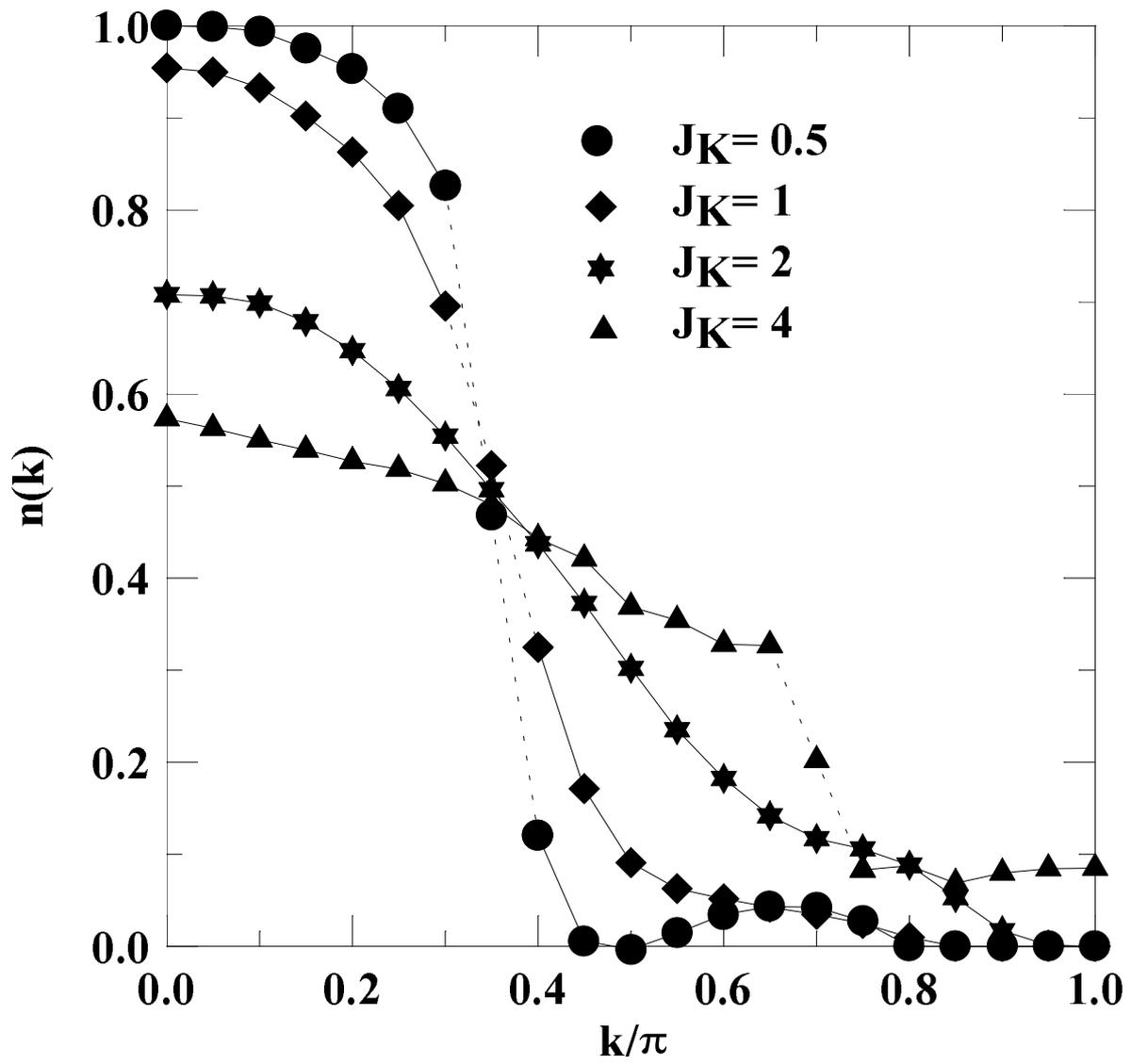
\begin{figure}[tbp]
\caption{ The electron momentum distribution $n(k)$ for $\rho =0.7$ at $%
J_K=0.5$ (circles), $J_K=1$ (diamonds), $J_K=2$ (stars) and $J_K=4$
(triangles).}
\end{figure}


\begin{references}
\bibitem{stewart}  G. R. Stewart, Rev. Mod. Phys. {\bf 56}, 755 (1984).

\bibitem{doniach}  S. Doniach, Physica {\bf 91B}, 231 (1977).

\bibitem{tsunetsugu1}  H. Tsunetsugu, Y. Hatsugai, K. Ueda and M. Sigrist,
Phys. Rev. B. 4{\bf 6}, 3175 (1992); C. C.\ Yu and White, Phys. Rev. Lett.
{\bf 71}, 3866 (1993); A. M. Tsvelik, Phys. Rev. Lett. {\bf 72}, 1048 (1994).

\bibitem{troyer}  M. Troyer and D. W\"urtz, Phys. Rev. B {\bf 47}, 2886
(1993).

\bibitem{tsunetsugu2}  H. Tsunetsugu,M. Sigrist and K. Ueda, Phys. Rev. B
{\bf 47}, 8345 (1993).

\bibitem{sigrist1}  M. Sigrist, H. Tsunetsugu, K. Ueda and T. M. Rice, Phys.
Rev. B {\bf 46}, 13838 (1992).

\bibitem{sigrist2}  M.\ Sigrist, H. Tsunetsugu and K. Ueda, Phys. Rev. Lett.
{\bf 67}, 2211 (1991).

\bibitem{white}  S. R. White, Phys. Rev. Lett. {\bf 69}, 2863 (1992).

\bibitem{chen}  Liang Chen and S.\ Moukouri, unpublished.

\bibitem{kondo}  J. Kondo, Phys. Rev. {\bf 154}, 644 (1967).

\bibitem{haldane}  F. D. M. Haldane, J. Phys. C {\bf 14}, 2585 (1981).

\bibitem{fazekas} P. Fazekas and E. M\"uller-Hartmann, Z. Phys. B {\bf 85},
285 (1991).


\bibitem{ueda} K. Ueda, T. Nishino and H. Tsunetsugu, Phys. Rev. B {\bf 50},
612 (1994).

\end{references}
\end{document}